\documentclass[aps,prl,floatfix,twocolumn,superscriptaddress, showpacs]{revtex4}
\usepackage{color}
\usepackage{bm}
\usepackage{amssymb}
\usepackage{amsmath}
\usepackage{amstext}
\usepackage{graphics}
\usepackage{epsfig}
\usepackage{placeins}
\usepackage{psfrag}
\usepackage{subfigure}

\begin{document}

\title{ DMRG Numerical Study of the Kagom\'{e} Antiferromagnet}
\author{H. C. Jiang}
\affiliation{Center for Advanced Study, Tsinghua University, Beijing, 100084, China}
\affiliation{Department of Physics and Astronomy, California State University,
Northridge, California 91330, USA}
\author{Z. Y. Weng}
\affiliation{Center for Advanced Study, Tsinghua University, Beijing, 100084, China}
\author{D. N. Sheng}
\affiliation{Department of Physics and Astronomy, California State University,
Northridge, California 91330, USA}
\date{April 10, 2008}

\begin{abstract}
We numerically study the spin-$\frac{1}{2}$ antiferromagnetic Heisenberg
model on the kagom\'{e} lattice using the density-matrix renormalization
group (DMRG)\ method. We find that the ground state is a magnetically
disordered spin liquid, characterized by an exponential decay of spin-spin
correlation function in real space and a magnetic structure factor showing
system-size independent peaks at commersurate antiferromangetic wavevectors.
We obtain a spin triplet excitation gap $\Delta E(S=1)=0.055\pm 0.005$ by
extrapolation based on the large size results, and confirm the presence of
gapless singlet excitations. The physical nature of such an exotic spin
liquid is also discussed.

\end{abstract}

\maketitle

\begin{figure}[tbp]
\centerline{
    \includegraphics[height=1.8in,width=3.0in]{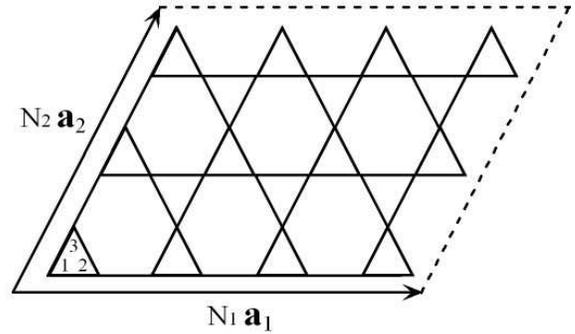}
    }
\caption{(color online) Sketch of a three-leg kagom\'{e} lattice with total
number of sites $N=3\times N_1 \times N_2$ and number of unit cells $%
N_{1}\times N_{2}=4\times 3$. Here $\mathbf{a}_{1}=(2,0)$ and $\mathbf{a}%
_{2}=(1,\protect\sqrt{3})$ are two primitive vectors of the unit cell
including three inequivalent sites (e.g., 1, 2, 3). }
\label{kagome_lattice}
\end{figure}

Novel magnetic properties and the possible existence of exotic spin liquid
states\cite{Anderson1973} in low-dimensional spin-$\frac{1}{2}$ systems have
attracted intensive attention in recent years. It has been established that
spins in the ground state of the two dimensional (2D) nearest-neighbor
Heisenberg antiferromagnet (HAF) model
\begin{equation*}
H=\sum_{\langle i,j\rangle }\mathbf{S}_{i}\cdot \mathbf{S}_{j}\text{ },
\end{equation*}%
are still ordered on square\cite{Huse1988} and triangular\cite%
{Bernu1992, Capriotti1999} lattice systems. However, spin
liquid states are likely to be found in some geometrically more frustrated
systems\cite{kagome}, like the kagom\'{e} lattice, which may be seen as a
diluted triangular lattice (see Fig. \ref{kagome_lattice}) with larger
geometrical frustration and lower coordination number than the triangular
lattice.  Earlier exact diagonalization (ED) studies \cite%
{Lecheminant1997,Waldtmann1998,Leung1993} suggest that the kagom\'{e}
antiferromagnet has a short-range spin correlation and a possible finite
spin gap $\sim 0.05$ when the finite-size results (up to $N=36$ sites) are
extrapolated to the thermodynamic limit. Within the spin gap, a large number
of singlet excited states are also identified.\cite{Leung1993,Mila1998}
Recently, algebraic vortex liquid  and Dirac spin liquid with \emph{gapless}
Dirac fermion excitations have been also proposed.\cite{matthew07, Ran2007}
Such a Dirac spin liquid state has a reasonably good variational energy,\cite%
{Ran2007} but the vanishing spin triplet gap is in contrast to the ED
result. While the discrepancy may be attributed\cite{Ran2007} to the
uncertainty of the finite size effect in the ED, alternatively a finite spin
gap can be also gained in the Dirac spin liquid state via an instability\cite%
{Hasting2001} towards a valence bond crystal (VBC) state with a broken
translational symmetry. Earlier on, Zeng and Marston\cite{Marston1991} also
proposed that the ground state of the kagom\'{e} HAF appears to be a VBC
state with a 36-site unit cell, which is supported by the series expansions.%
\cite{Singh2007} \textbf{\ }So far\textbf{\ }the precise nature of the HAF
on the kagom\'{e} lattice in the long-wavelength and low-energy regime
remains unsettled.

Experimentally the newly synthesized Herbertsmithite $\mathrm{%
ZnCu_{3}(OH)_{6}C_{l2}}$ has brought tremendous excitement to this field, in
which the spin-$\frac{1}{2}$ copper ions form layered kagom\'{e} lattices.
The absence of the magnetic ordering has been established based on the
neutron scattering measurement\cite{Helton2007} down to $50mK$, as compared
to a relatively high Curie-Weiss temperature ($\sim 300K$). The magnetic
measurements\cite{Ofer2006, Mendels2007, Helton2007, Alariu2008} also
suggest that there is no signature of a finite spin gap seen in the
experiment, which seems consistent with an algebraic spin liquid, but
contrary to a short-range spin liquid state with a finite triplet gap.
However, possible impurity spins outside the kagom\'{e} layers, caused by
substitutions of nonmagnetic Zn sites with Cu, or the presence of
Dzyaloshinsky-Moriya (DM) interactions,\cite{Rigol2007} may all play an
important role in order to fully understand the experimental results. While
the experimental situation is still unclear, on the fundamental side, it is
highly desirable to reexamine the issues regarding the nature of the ground
state and low-lying excitations in a pure spin-$\frac{1}{2}$ HAF model on
the kagom\'{e} lattice.

Due to the geometrical frustration of the kagom\'{e} lattice, the quantum
Monte Carlo (QMC) method encounters the sign problem, whereas the ED
calculation is restricted to small system size. In this Letter, we present a
systematic numerical study by employing the DMRG method\cite{White1992},
with keeping a large number of basis states in the DMRG blocks. We find that
the ground state is indeed a magnetically disordered state, which is
characterized by an exponential decay of the equal-time spin-spin
correlation function in real space. The corresponding magnetic structure
factor shows small peaks at commensurate momenta, with near constant peak
values insensitive to the size of the system, in sharp contrast to the
structure factor of the magnetic ordered state on a triangular lattice.
Furthermore, we calculate the spin triplet gap, which is extrapolated to a
finite value $\Delta E(S=1)=0.055\pm 0.05$ in the large sample size limit.
In this spin liquid state, there also exist low-lying singlet excitations,
with their gap approaching zero at large sample size limit. Our calculations
strongly hint that the ground state may be described by a resonating valence
bond (RVB) spin liquid with short-range antiferromagnetic correlations and
weak bond-bond and chirality-chirality correlations without explicitly
breaking translational and rotational symmetries.

We consider a kagom\'{e} lattice with finite length vectors $N_{1}\mathbf{a}%
_{1}$ and $N_{2}\mathbf{a}_{2}$ as shown in Fig. \ref{kagome_lattice}. Here $%
\mathbf{a}_{1}=(2,0)$ and $\mathbf{a}_{2}=(1,\sqrt{3})$ are two primitive
vectors of the unit cell which includes three lattice sites on a triangle.
The total number of sites is $N=3\times N_{1}\times N_{2}$, with the number
of unit cells $N_{1}\times N_{2}$. We will extend the calculation from $N=36$
(the maximum size for ED) to much larger sizes with different geometries, up
to $N=3\times 16\times 4$ ($192$ sites), using the DMRG method. To test the
performance of the DMRG method in the 2D spin systems, we have compared our
results with the ED up to $N=36$ sites for various lattices (including
triangular, square, and kagom\'{e} lattices) and obtained accurate ground
state energies with errors smaller than 0.01\%. For present study, we keep
up to $m=4096$ states in the DMRG block for most systems with more than $24$
sweeps to get a converged result, and the truncation error is of the order
or less than $10^{-5}$. We make use of the periodic boundary condition (PBC)
to reduce the finite-size effect for a more reliable extrapolation to the
thermodynamic limit.

\begin{figure}[tbp]
\centerline{
    \includegraphics[height=3.6in,width=3.6in]{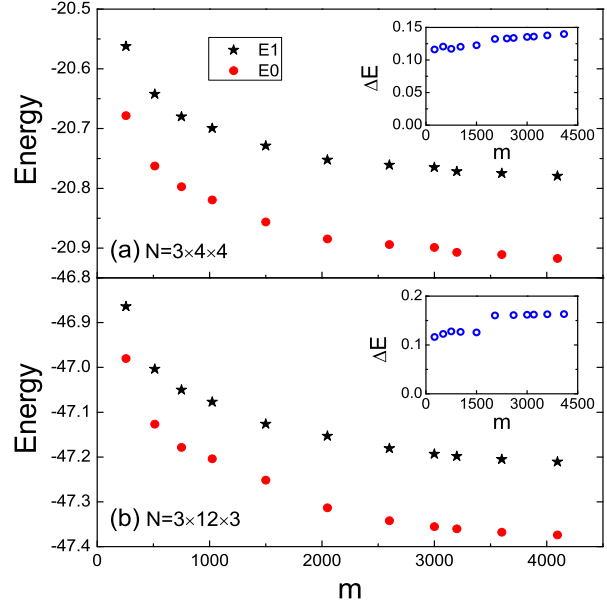}
    }
\caption{(color online) The ground state energy $E_{0}$ (solid circles) and
the excitation energy $E_{1}$ (solid stars) in the total spin $S=1$ sector
are shown as a function of $\ m$ (the number of states kept in each block)
for the system with $N=3\times 4\times 4$ in (a) and $N=3\times 12\times 3$
in (b). The energy gap for spin-1 excitation gap $\Delta E=E_{1}-E_{0}$
(open circles) as a function of $m$ for these two systems are shown in the
insets. }
\label{evolution}
\end{figure}

We first present the DMRG result for a system with $N=48$ sites ($%
N_{1}=N_{2}=4$). In Fig. \ref{evolution}(a), we show the ground state energy
$E_{0}$ as a function of $m$ --- the number of states kept in each block
(the dimension of the Hilbert space $=4m^{2}$). The ground state energy is
extrapolated to $-20.958$ at large $m$ limit and the estimated error at $%
m=4096$ is about $0.16\%$. Similarly, the lowest energy $E_{1}$ in
the total spin $S=1$ sector is also shown in Fig.
\ref{evolution}(a). Define the spin triplet gap $\Delta E(S=1)\equiv
E_{1}(S=1)-E_{0}$. As plotted in the inset of Fig.
\ref{evolution}(a), such a spin gap starts to saturate at $m>2000$
and approaches the value $0.145$ at large $m$. In Fig.
\ref{evolution}(b), similar results for a larger system with $N=108$
($3\times 12\times 3$) are also shown, where we find a slightly
larger spin gap at $0.163$ for this larger but narrow system.

\begin{figure}[tbp]
\centerline{
    \includegraphics[height=3.0in,width=3.8in]{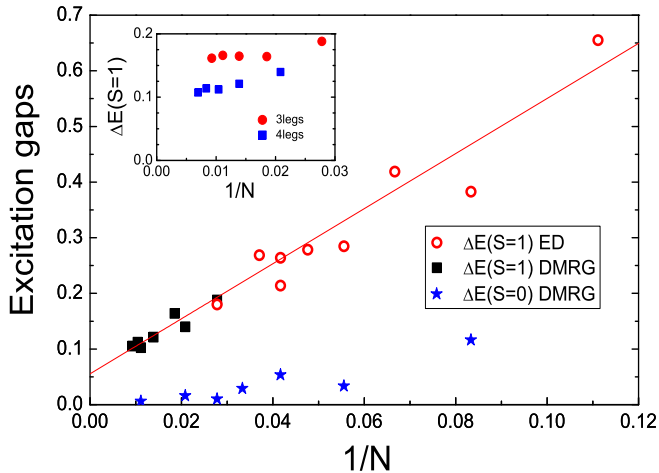}
    }
\caption{(color online) The spin gap $\Delta E(S=1)$ for square-like systems
(see text) at different system sizes obtained from the ED\protect\cite%
{Waldtmann1998} (open circles) and DMRG (solid squares) with $N=36-108$. The
singlet excitation gap $\Delta E(S=0)=E_{1}(S=0)-E_{0}$ is also given (solid
stars). Inset shows the spin gap for both 3-leg and 4-leg systems.}
\label{spin_gap}
\end{figure}

A systematic size dependence of the spin gap is shown in Fig. \ref{spin_gap}%
. In the main panel, the spin gap $\Delta E(S=1)$ (solid squares) vs. $1/N$
is plotted with $N=3\times 4\times 3$, $3\times 6\times 3$, $3\times 4\times
4$, $3\times 6\times 4$, $3\times 8\times 4$, $3\times 6\times 5$ and up to $%
3\times 6\times 6=108$, together with the results of the ED\cite%
{Waldtmann1998} (open circles) at smaller sizes (note that $N=36$ site
system in the ED has a different geometry as compared to the $N=3\times
4\times 3$ system in the present calculation). All these data follow nicely
a straight line shown in Fig. \ref{spin_gap}, which allows us to extrapolate
the spin gap to a finite value $\Delta E(S=1)=0.055\pm 0.005$ in the
thermodynamic limit. Note that all the data presented in the main panel are
for the systems close to square-like with the aspect ratio $\alpha
=N_{1}/N_{2}$ in the range of $1\leq \alpha \leq 2.$ The corresponding
ground state energies per site $\epsilon _{0}$ and the spin gaps $\Delta
E(S=1)$ for the various system sizes at a given $m=4096$ are listed in Table
I. Furthermore, in the inset of Fig. \ref{spin_gap}, the spin gap vs. $1/N$
for 3-leg ($N=3\times N_{1}\times 3$) and 4-leg ($N=3\times N_{1}\times 4$)
systems with $N_{1}=4-12$ (thus including larger $\alpha $'s) are also
present for comparison. In general, the spin gap of the 4-leg systems is
smaller than that of the 3-leg systems due to the finite-size effect,
consistent with the behavior shown in the main panel for the more
square-like systems.

Besides the spin triplet gap $\Delta E(S=1)$, the lowest singlet excitation
energy $\Delta E(S=0)\equiv E_{1}(S=0)-E_{0}$ is also shown in the main
panel of Fig. \ref{spin_gap} (solid stars), whose magnitude is much smaller
than $\Delta E(S=1)$ and approaches zero with increasing sample size. This
is consistent with the ED results\cite{Leung1993,Mila1998} at smaller
systems, in which a large number of singlet states below the spin gap,
growing with the system size, are identified. In contrast to the finite spin
triplet gap, such vanishing singlet excitation energy indicates that the
low-lying singlet excitations will play a dominant role in the
low-temperature thermodynamic properties like specific heat.

\begin{table}[h]
\caption{The ground state energy per site $\protect\epsilon _{0}$ and spin
gap $\Delta E(S=1)$ for the square-like kagom\'{e} lattice, obtained by the
DMRG with keeping $m=4096$\protect\cite{Cal_detail} basis states in one
block.}
\label{table:spin_gap_square}\centering \vspace{2mm}
\begin{tabular}[b]{|c|c|c|}
\hline
\hspace{10mm} $N$\hspace{10mm} & \hspace{8mm} $\epsilon _{0}$\hspace{8mm} &
\hspace{6mm} $\Delta E(S=1)$\hspace{6mm} \\ \hline
$3\times 4\times 3$ & -0.43898 & 0.188 \\ \hline
$3\times 6\times 3$ & -0.43875 & 0.164 \\ \hline
$3\times 8\times 3$ & -0.43867 & 0.165 \\ \hline
$3\times 10\times 3$ & -0.43868 & 0.163 \\ \hline
$3\times 12\times 3$ & -0.43865 & 0.163 \\ \hline
$3\times 4\times 4$ & -0.43591 & 0.140 \\ \hline
$3\times 6\times 4$ & -0.43564 & 0.122 \\ \hline
$3\times 8\times 4$ & -0.43556 & 0.112 \\ \hline
$3\times 10\times 4$ & -0.43552 & 0.114 \\ \hline
$3\times 6\times 6$ & -0.43111 & 0.105 \\ \hline
\end{tabular}%
\end{table}

\begin{figure}[tbp]
\centerline{
    \includegraphics[height=3.6in,width=3.6in]{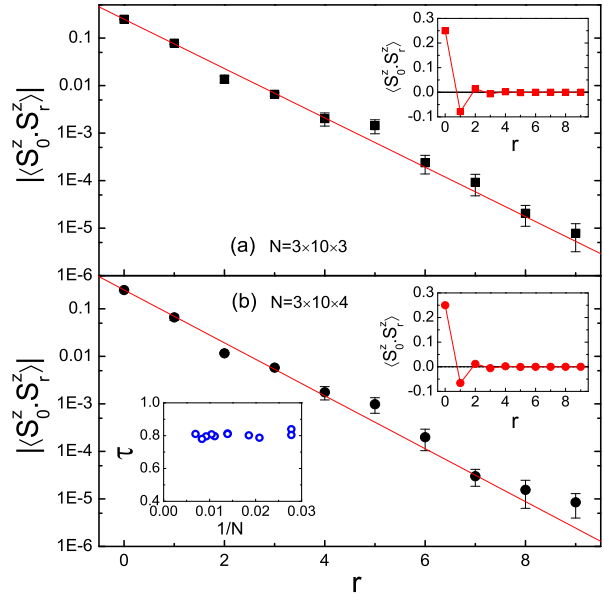}
    }
\caption{(color online) The spin-spin correlations $|\langle
S_{0}^{z}S_{r}^{z}\rangle |$ along the $\mathbf{a}_{1}$ direction with $%
N=3\times 10\times 3$ and $N=3\times 10\times 4$. The error bar represents
the mean square deviation of all the equivalent sites. The straight line is
a fitting to an exponential function $|\langle S_{0}^{z}S_{r}^{z}\rangle
|=A\exp (-r/\protect\tau )$. The system size dependence of the correlation
length $\protect\tau $ is shown in the lower left inset. The spin-spin
correlations $\langle S_{0}^{z}S_{r}^{z}\rangle $ are also given in the
insets. }
\label{spin_cor}
\end{figure}

To characterize the ground state, the spin-spin correlation function $%
|\langle S_{0}^{z}S_{r}^{z}\rangle |$ is presented in Fig. \ref{spin_cor}
for two systems with $N=3\times 10\times 3$ and $3\times 10\times 4$,
respectively. Here $r$ is the distance between the two sites along the $%
\mathbf{a}_{1}$ direction in units of the lattice constant and the error bar
denotes the mean square deviation for all the equivalent pairs of sites.
Fig. \ref{spin_cor} shows that the results are well fitted by the straight
lines representing an exponential fit: $|\langle S_{0}^{z}S_{r}^{z}\rangle
|=A\exp (-r/\tau )$ with $\tau $ as the spin correlation length whose size
is insensitive to the number of legs and is about $0.8$ lattice spacing for
both systems (In the left lower inset of Fig.\ref{spin_cor}(b), $\tau $ as a
function of $N$ is shown for a few systems up to $N=144)$. These results
clearly illustrate that the ground state is magnetically disordered with no
long-range correlations. Furthermore, $\langle S_{0}^{z}S_{r}^{z}\rangle $
itself exhibits short-range antiferromagnetic oscillations commensurate with
the lattice constant along the $\mathbf{a}_{1}$ direction as shown in the
top right insets of Fig. \ref{spin_cor}. Completely similar results are
found for the transverse spin-spin correlation $\langle
S_{i}^{+}S_{j}^{-}\rangle $ due to the spin rotational symmetry.

\begin{figure}[tbp]
\centerline{
    \includegraphics[height=3.2in,width=3.6in]{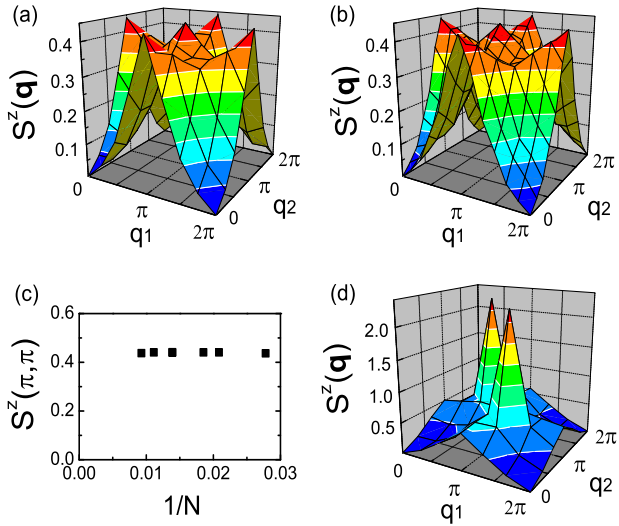}
    }
\caption{(color online) The static structure factors for the kagom\'{e}
systems with $N=3\times 4\times 4$ in (a); $3\times 6\times 6$ in (b); The
peak values $S^{z}(\protect\pi ,\protect\pi )$ vs. $1/N$ are plotted in (c);
In (d), the structure factor $S^{z}(\mathbf{q})$ for a triangular HAF
$N=6\times 6$ system is also shown for comparison. }
\label{struc_factor}
\end{figure}

To further describe the short-range magnetic correlations, we calculate the
static structure factor $S^{z}(\mathbf{q})={\frac{1}{N}}\sum_{ij}e^{i\mathbf{%
q}\cdot (\mathbf{r}_{i}-\mathbf{r}_{j})}\langle S_{i}^{z}S_{j}^{z}\rangle $
and present the results in Fig. \ref{struc_factor} (a) and (b) for system
sizes $N=3\times 4\times 4$ and $N=3\times 6\times 6$, respectively, where $%
\mathbf{q}$ are allowed magnetic wavevectors with components $%
(q_{1},q_{2})=2\pi (n_{1}/N_{1},n_{2}/N_{2})$ along the directions of two
primitive basis vectors in the reciprocal lattice. These figures show that $%
S^{z}(\mathbf{q})$ exhibits small peaks at $q^{\ast }=(\pi ,0)$, $(0,\pi )$
and $(\pi ,\pi )$, indicating the dominant short-range antiferromagnetic
correlations between the nearest neighbor sites. It is important to observe
that the peaks remain at the same small value $\sim 0.44$ without changing
much with increasing system size, as clearly illustrated by Fig. \ref%
{struc_factor} (c) for the peak values at $q^{\ast }=(\pi ,\pi )$. Such weak
and size-independent peaks are in sharp contrast to the structure factor of
a magnetic ordered system in the triangular HAF model as shown in Fig. \ref%
{struc_factor} (d) with $N=6\times 6$ sites, where sharp peaks appear at $%
q^{\ast }=(\frac{4\pi }{3},\frac{2\pi }{3})$ and $(\frac{2\pi }{3},\frac{%
4\pi }{3})$.

Finally, we have also checked the bond-bond and chirality-chirality
correlations and found that both are short-ranged with exponential-decay
behavior. Thus, there seems no explicitly broken translational or rotational
symmetry to be responsible for the gapless singlet excitation found in such
a system. However, for our DMRG calculation with finite system size (up to
192 sites), it is difficult to detect the possible VBC ordering with
extremely large unit cell of 36-sites.\cite{Marston1991, Singh2007} On the
other hand, the overall features of the structure factor in Fig. \ref%
{struc_factor} (a) and (b) are quite similar to those calculated\cite{tli}
based on the Gutzwiller projected Dirac spin liquid state,\cite{Ran2007}
although the spin gap vanishes in the latter. One may thus conjecture the
ground state for the kagom\'{e} HAF be described by an RVB state, which is
similar to the projected Dirac spin liquid state at short ranges. But at
long ranges it will have a finite spin triplet gap because of the finite
size of spin RVB pairing, and the gapless singlet excitations are Goldstone
modes originated from the broken $U(1)$ gauge symmetry due to the RVB
condensation, like in a charge-neutral superconductor. We shall use the
variational QMC method to further study such kind of spin liquid states and
compare with the numerical results elsewhere.

In summary, we have numerically studied the ground state properties and
low-lying excitations of the kagom\'{e} antiferromagnet using the DMRG
method. Our results provide strong evidence that the ground state is a spin
liquid with only short-range antiferromagnetic correlations without a
magnetic order or other translational or rotational symmetry breaking. The
spin triplet excitation has a gap extrapolated to a finite value in the
thermodynamic limit, but the singlet excitation remains gapless. The nature
of such a spin liquid state has been discussed based on the numerical
results.

\textbf{Acknowledgment:} We are grateful for stimulating discussions with L.
Balents, Z. C. Gu, T. Li, T. Xiang, and Y. Yue. This work is supported by
the NSFC grant no. 10688401 (HCJ, ZYW), the DOE grant DE-FG02-06ER46305, the
NSF grants DMR-0605696 and DMR-0611562 (HCJ, DNS).


\begin{thebibliography}{99}
\bibitem{Anderson1973} P. W. Anderson, Mater. Res. Bull, \textbf{8}, 153
(1973).

\bibitem{Huse1988} D. A. Huse and V. Elser, Phys. Rev. Lett. \textbf{60},
2531 (1988).


\bibitem{Bernu1992} B. Bernu, \emph{et al.}, Phys. Rev. Lett. \textbf{69},
2590 (1992).

\bibitem{Capriotti1999} L. Capriotti, \emph{et al.}, Phys. Rev. Lett.
\textbf{82}, 3899 (1999).


\bibitem{kagome} F. Wang and  A. Vishwanath, Phys. Rev. B {\textbf 74},
174423 (2006).

\bibitem{Lecheminant1997} P. Lecheminant $et$ $al$., Phys. Rev. B \textbf{56}%
, 2521 (1997).

\bibitem{Waldtmann1998} Ch. Waldtmann $et$ $al$., Eur. Phys. J. B \textbf{2,
}501 (1998).

\bibitem{Leung1993} P. W. Leung, and Veit Elser, Phys. Rev. B \textbf{47},
5459 (1993).

\bibitem{Mila1998} F. Mila, Phys. Rev. Lett. \textbf{81}, 2356 (1998).

\bibitem{matthew07} S. Ryu, O. I. Motrunich, J. Alicea, Matthew P. A.
Fisher, Phys. Rev. B \textbf{75}, 184406 (2007).

\bibitem{Ran2007} Y. Ran $et$ $al$., Phys. Rev. Lett. \textbf{98}, 117205
(2007).

\bibitem{Hasting2001} M. B. Hastings, Phys. Rev. B \textbf{63}, 014413
(2001).

\bibitem{Marston1991} J. B. Marston and C. Zeng, J. Appl. Phys. \textbf{69},
5962 (1991).

\bibitem{Singh2007} R. R. P. Singh, David A. Huse Phys. Rev. B \textbf{76},
180407 (2007).

\bibitem{Helton2007} J. S. Helton $et$ $al$., Phys. Rev. Lett. \textbf{98},
107204 (2007).

\bibitem{Ofer2006} O. Ofer $et$ $al$. , cond-mat/0610540.

\bibitem{Mendels2007} P. Mendels $et$ $al$., Phys. Rev. Lett. \textbf{98},
077204 (2007).

\bibitem{Alariu2008} A. Olariu $et$ $al$., Phys. Rev. Lett. \textbf{100},
087202 (2008).

\bibitem{Rigol2007} M. Rigol and R. R. P. Singh, Phys. Rev. Lett. \textbf{98}%
, 207204 (2007).

\bibitem{White1992} Steven R.White, Phys. Rev. Lett. \textbf{69}, 2863
(1992); \textit{ibid.} \textbf{77}, 3633 (1996).

\bibitem{Cal_detail} For example, for $N=3\times 6\times 6$, the energy $%
\epsilon _{0}=-0.43111$ at $m=4096$ is shown in Table I. We have also
calculated $\epsilon _{0}$ as a function of $m$ (up to $m=4096)$, which give
rise to an extrapolation of the energy $\epsilon _{0}=-0.43308$ at $%
m\rightarrow \infty $.

\bibitem{tli} T. Li, unpublished.
\end{thebibliography}
\end{document}